# Split Fermi Surfaces of the Spin–Orbit-Coupled Metal $Cd_2Re_2O_7$ Probed by de Haas–van Alphen Effect

Yasuhito Matsubayashi[1], Kaori Sugii[1], Hishiro T. Hirose[2], Daigorou Hirai[1], Shiori Sugiura[2], Taichi Terashima[2], Shinya Uji[2], and Zenji Hiroi[1*]

[1]*Institute for Solid State Physics, University of Tokyo, Kashiwa, Chiba 277-8581, Japan*

[2]*National Institute for Materials Science, Tsukuba, Ibaraki 305-0003, Japan*

The superconducting pyrochlore oxide $Cd_2Re_2O_7$ shows a structural transition with inversion symmetry breaking (ISB) at $T_{s1}$ = 200 K. A recent theory [L. Fu, Phys. Rev. Lett. **115**, 026401 (2015)] suggests that the origin is an electronic instability that leads to a multipolar order in the spin–orbit-coupled metal. To observe the Fermi surface of the low-temperature phase of $Cd_2Re_2O_7$, we perform de Haas–van Alphen effect measurements by means of magnetic torque. In reference to a calculated band structure, the spin-split Fermi surfaces with large cyclotron masses of $5-9m_0$ are revealed. The splitting is suggested to be due to an antisymmetric spin–orbit coupling induced by ISB, the strength of which is estimated to be approximately 67 K, which is rather smaller than those of typical non-centrosymmetric metals.

$Cd_2Re_2O_7$ is the only superconductor ($T_c$ = 1.0 K) among many pyrochlore oxides with the chemical formula $A_2B_2O_7$.[1-4] Upon cooling, it shows two structural phase transitions from $Fd\bar{3}m$ (phase I) to $I\bar{4}m2$ (phase II) at $T_{s1}$ = 200 K, and then to $I4_122$ (phase III) at $T_{s2}$ = 120 K.[5] Inversion symmetry breaking (ISB) occurs at the first transition. The origin of the first transition is suggested to be the band Jahn–Teller effect that lifts the degeneracy of the Fermi surface, because a large decrease in the density-of-states is observed by magnetic susceptibility and nuclear magnetic resonance measurements below $T_{s1}$.[6-8]

A recent theory by L. Fu suggests that the ISB transition of $Cd_2Re_2O_7$ is driven by a Fermi liquid instability for the spin–orbit-coupled metal (SOCM), which is a metal with a centrosymmetric crystal structure and a strong spin–orbit coupling.[9] The SOCM possesses three kinds of Fermi liquid instabilities induced by spin–orbit coupling toward gyrotropic, multipolar, and ferroelectric orders in terms of symmetry. The irreducible representation of each order has odd parity and can be coupled with an odd-parity phonon at the Γ point, which results in a parity-breaking transition. The soft phonon of the successive transitions of $Cd_2Re_2O_7$ is a $E_u$ mode in the $O_h$ point group,[10,11] which coincides with the irreducible representation of the multipolar order. Therefore, $Cd_2Re_2O_7$ is considered to be an SOCM candidate with a multipolar order.[9] However, the recent second harmonic generation study suggests that the order parameter of the ISB transition belongs to a $T_{2u}$ multipolar order, which eventually induces a $E_u$ distortion.[12-14] Thus, the mechanism of the ISB transition of $Cd_2Re_2O_7$ is still under debate.

In noncentrosymmetric metals, the spin splitting of Fermi surface occurs owing to antisymmetric spin–orbit coupling (ASOC), which has been intensively studied in the field of spintronics.[15] Moreover, exotic superconductivity with the mixing of *s*- and *p*-wave symmetries can occur in the absence of inversion symmetry,[16] which appears to be the case for $Cd_2Re_2O_7$.[17-19] Spin splitting due to ASOC has been experimentally observed by means of the de Haas–van Alphen (dHvA) oscillation[20,21] and spin-/angle-resolved photoemission spectroscopy.[22] The ISB transition of $Cd_2Re_2O_7$ should involve the deformation and spin-splitting of the Fermi surface but has not yet been directly observed; previous transport measurements suggested certain changes in the Fermi surface structure at $T_{s1}$ as well as at $T_{s2}$.[4] In this study, to clarify the electronic states of $Cd_2Re_2O_7$ and to obtain insight into a possible multipolar order, we perform dHvA oscillation measurements by means of magnetic torque on high-quality crystals.

Crystals of $Cd_2Re_2O_7$ were grown by a chemical vapor transport technique, the details of which have been reported elsewhere.[23] The residual-resistivity ratio (RRR), defined as $\rho$ (300 K)/$\rho$ (3 K) in the present study, was at most 40 in previous reports,[4] while we have obtained higher-quality crystals with RRR = 100–300 through the improvements in synthesis method and condition. For the dHvA experiments, two octahedral crystals of size 0.3 mm were used, which had been selected from two batches with average RRRs of 190 and 40. The dHvA measurements at temperatures down to 30 mK and in magnetic fields up to 17.5 T were conducted at the Tsukuba Magnet Laboratory, National Institute for Materials Science. dHvA oscillations were detected by magnetic torque measurements on a crystal attached to a microcantilever. The two rotation axes of the magnetic field employed were [1–10] (rotation A) and [11–2] (rotation B); for clarity, the orientation in cubic phase I (fcc) is used instead of that in tetragonal phase III (bct). Field angle $\theta$ is defined such that $\theta$ = 180° corresponds to $B$ // [111].

To interpret the observed dHvA oscillations, the electronic structure of $Cd_2Re_2O_7$ is investigated by fully relativistic calculations including the effect of spin–orbit coupling using the full-potential linearized augmented plane wave method implemented in the WIEN2k package.[24] The

Perdew–Burke–Ernzerhof parameterized generalized gradient approximation[25]) is employed for the exchange correlation potential. The calculation is performed based on the crystal structure of phase III ($I4_122$) at 90 K reported in the reference.[26]) The irreducible Brillouin zone is sampled in a 10 × 10 × 10 k-mesh for self-consistent field calculation, and 30 × 30 × 30 k-mesh for calculation of the Fermi surfaces. Simulations of dHvA frequencies are performed using the SKEAF code.[27])

Figure 1(a) shows the field-dependencies of the magnetic torque of a $Cd_2Re_2O_7$ crystal with RRR = 190 at 30 mK and at $B$ // [001] and [111] after the subtraction of background signals proportional to $B^2$. Complex dHvA oscillations are clearly observed in each case. Figure 1(b) shows the Fourier transform spectra, in which many dHvA peaks are observed at frequencies up to 2250 T, giving the extremal cross-sectional areas of the Fermi surfaces. The branches of the dHvA signals at $B$ // [111] are assigned from α to κ with increasing frequency. It is noteworthy that three kinds of domains are formed at the $T_{s1}$ transition from cubic to tetragonal, such that the dHvA signals from all the three domains are included in the observed signals. However, those branches assigned at $B$ // [111] should represent independent cross sections, because the three domains become equivalent at this field direction. The angular dependencies of the dHvA frequency are shown in Figs. 2(a) and (b) for the rotation axes along [1–10] and [11–2], respectively. Many branches are clearly observed up to high frequencies of 4000 T.

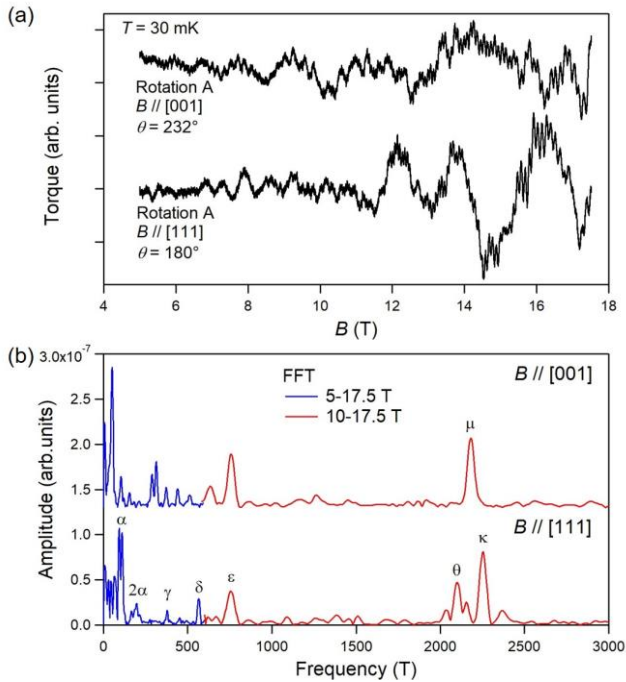

**Fig. 1.** (Color online) (a) Field dependencies of magnetic torque for a $Cd_2Re_2O_7$ crystal with RRR = 190 at $T$ = 30 mK and $B$ // [001] and [111] in the cubic notation of phase I. The rotation axis is [1–10] (rotation A). The torque data for $B$ // [001] was measured at $\theta$ = 232°, which is titled by 3° from the exact angle of $\theta$ = 235°, to enhance the intensity. (b) Fourier transform spectra of the dHvA oscillations. The field ranges for Fourier transformation are set to 5–17.5 T for frequencies lower than 600 T, and to 10–17.5 T for higher frequencies. The observed peaks at $B$ // [111], where three kinds of the tetragonal domains become equivalent, are assigned as α, γ, δ, ε, θ, κ in sequence, and that at 2180 T for $B$ // [001] is named as μ.

For another crystal with RRR = 40, however, similar dHvA oscillations were observed, but their amplitudes were small with some of the branches observed for the crystal with RRR = 190 missing. This fact clearly indicates that the crystal quality is lower for RRR = 40. The Dingle temperatures estimated for the κ branch at $B$ // [111] are 0.40 K and 1.3 K, and the mean free paths are 117 nm and 36 nm for the crystals with RRR = 190 and 40, respectively. Thus, the RRR is a good measure of the crystal quality in $Cd_2Re_2O_7$, which is not always the case for a metal with a phase transition. Herein, we analyze the clear dHvA signals from the RRR = 190 crystal and discuss on the Fermi surfaces of $Cd_2Re_2O_7$.

To assign the observed branches, band structure calculation has been performed for phase III. Phase I is a compensated metal with two round electron-like Fermi surfaces around the Γ point and one hole surface near the K point.[8, 28]) Nevertheless, the previous band structure calculation for phase III[26]) suggests the existence of seven Fermi surfaces including two-dimensional cylindrical electron surfaces, which are completely different from those of phase I. However, the calculation assumed strong electron correlations even though most previous experiments indicated weak correlations.[4]) Thus, we perform an alternative calculation with weak electron correlations for phase III and investigate how the Fermi surfaces of phase I change in phase III by ISB. It is naively expected that the phase III activated ASOC induces spin splittings for the three Fermi surfaces of phase I, resulting in six Fermi surfaces. In fact, our calculation, as shown in Fig. 3, reveals three pairs of spin-split Fermi surfaces: #271 and #272 (hole), #273 and #274 (electron), and #275 and #276 (electron). Three-dimensional spin splitting of Fermi surfaces around the Γ point is obvious particularly for the pair of electron surfaces, #275 and #276, which reflects the symmetry of the $g$ vector that characterizes ASOC for the space group $I4_122$: $(\alpha_1(k_x s_x + k_y s_y) + \alpha_2 k_z s_z)$.[29])

The calculated dHvA frequencies are compared with the experiments in Fig. 2. The largest Fermi surface of hole #272 produces many dispersive branches, while the pairs of electron Fermi surfaces produce nearly flat branches. The intense branch increasing from 760 T at $\theta$ = 235° to 2250 T at $\theta$ = 180° (θ or κ branch) in rotation A may correspond to the #272 hole band. In addition, the highly dispersive branch rising up around $B$ // [111] in rotation A and those crossing near 3000 T at $\theta$ = 121° in rotation B (ν branch) may also come from this Fermi surface. Thus, our calculation can reproduce the experiments to some extent. However, the simulations indicate that the counterpart #271 hole surface shrinks and has low-lying branches near 100 T. Although the corresponding signals appear in Fig. 2, the identification of these branches is difficult because tiny pockets from other Fermi surfaces may appear at this low frequency range. It is also likely that the #271 hole surface has been completely removed but appears in the calculation within the accuracy.

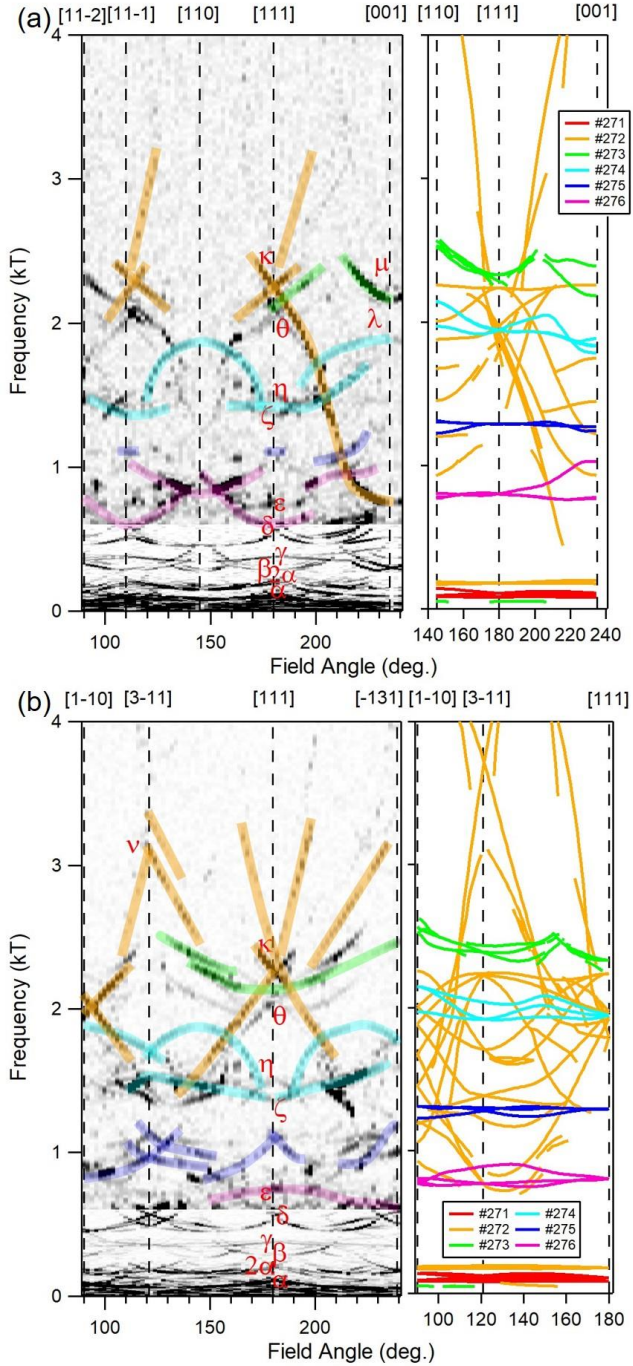

**Fig. 2.** (Color online) Angular dependences of dHvA frequencies with rotation axes along (a) [1–10] (rotation A) and (b) [11–2] (rotation B). The angle of the applied magnetic field is defined as 180° at $B$ // [111]. The simulated angle dependencies of the dHvA frequencies from all six Fermi surfaces shown in Fig. 3 are plotted in the right panels with the corresponding colors. Contributions from three kinds of tetragonal domains are considered, which happen to merge at $B$ // [111]. The several distinct experimental branches shaded by colors may correspond to those with the same colors in the simulations. In addition to the branches around $B$ // [111] ($\alpha$ to $\kappa$), two branches near $B$ // [001] in rotation A and one near $B$ // [3–11] in rotation B are assigned as $\lambda$, $\mu$, and $\nu$, respectively.

In addition, the round electron surfaces are expected to generate flat branches. In fact, the branches from the pair of #273 and #274 Fermi surfaces lie at 2200–2600 T and 1800–2100 T, respectively. The former may correspond to the intense branch at 2200 T near $B$ // [001] in rotation A ($\mu$ branch), while the latter to the weak branch at 1900 T ($\lambda$ branch). The #275 and #276 electron surfaces are also expected to produce flat branches lying around 1200–1300 T and 800–1000 T, respectively, which may correspond to the flat branches lying near 1000–1300 T and 600–1000 T ($\varepsilon$ or $\delta$), respectively. Note that the experimental branches appear more dispersive than those in the simulations, suggesting that the actual electron Fermi surfaces are more deformed.

Our band structure calculations can approximately reproduce the experimental results, given some deviations of the absolute value of the dHvA frequency. However, significant mismatches remain, which may be due to the accuracy and reliability of our band structure calculation. One problem comes from the uncertainty of the employed crystal structure parameters, especially the positions of the oxygen atoms.[26] A neutron diffraction experiment using high-quality crystals to obtain reliable structure parameters for all the three phases and more accurate band-structure calculations are in progress.

Figure 4 shows the temperature dependencies of the amplitudes of the dHvA signals at three frequencies: 2200 T at $B$ // [001] and 2250 T at $B$ // [111] in rotation A and 3080 T at $\theta = 232°$ in rotation B. The first signal may be derived from the electron surfaces, and the second and third from the hole surfaces. The cyclotron mass of carrier $m^*$ is estimated by fitting the temperature damping $R_T$ to the Lifshitz–Kosevich formula:[30]

$$R_T = \frac{Kp\mu T/H}{\sinh(Kp\mu T/H)}, \quad (1)$$

where $K = 14.7$ T K$^{-1}$, $p$ is the order of harmonics, $\mu = m^*/m_0$ ($m_0$: mass of free electron), $T$ is the temperature, and $H$ is the magnetic field. The obtained cyclotron masses are $5.5m_0$, $7.8m_0$, and $9.3m_0$, and the corresponding band masses $m_{band}$ are $1.75m_0$, $3.14m_0$, and $6.90m_0$ for the three bands, respectively. Thus, both the electron and hole bands with frequencies over 2000 T have large $m^*/m_{band}$ values of 1.35–3.14. These results are consistent with the fact that the Sommerfeld coefficient from the heat capacity is more than three times larger than the bare band mass.[4,6] The heavy electron states are possibly induced by some kind of many-body effects. The electron–phonon interaction must be weak in $Cd_2Re_2O_7$ because the superconducting state is of the weak-coupling BCS type, and the electron correlation is also weak from the small Wilson ratio.[4-6] Therefore, ASOC and fluctuations of multipolar order are likely to play an important role as many-body effects leading to the mass enhancement.[4]

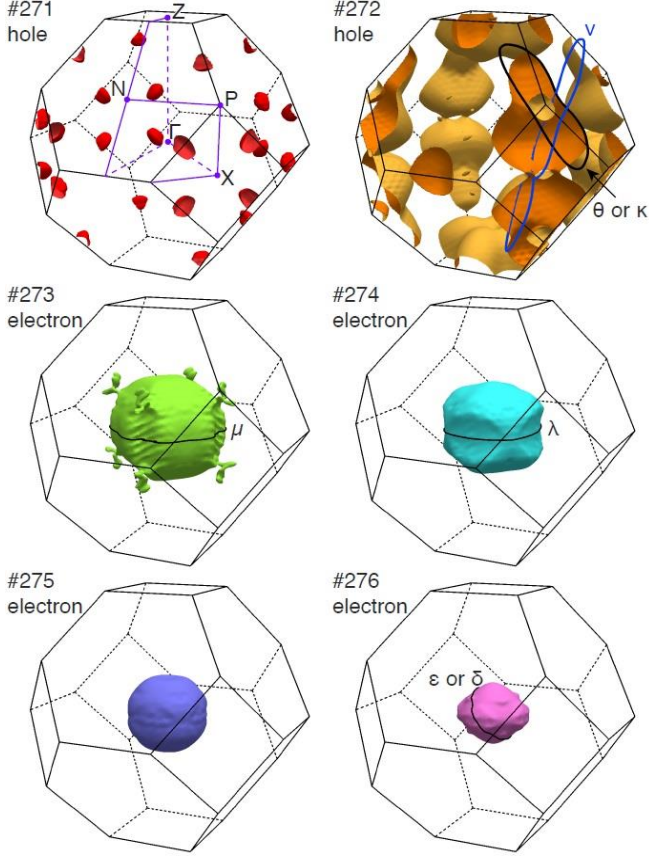

**Fig. 3.** (Color online) Fermi surfaces calculated for phase III and the extremal cross-sections, which have possibly been observed experimentally. The structural parameters from ref. 26 are employed for the calculations. The hole surface around the K point in phase I splits into #271 and #272, and the two electron surfaces around the Γ point split into #273/#274 and #275/#276 in phase III.

Here, we make a crude evaluation of the magnitude of ASOC in $Cd_2Re_2O_7$ based on the observed spin splitting of the electron Fermi surfaces around the Γ point. In the case of noncentrosymmetric metals with the Rashba-type spin–orbit coupling, the magnitude of the transverse component of ASOC in the $k_x$–$k_y$ plane is given by $2|\alpha^* p_\perp|$, where $\alpha^*$ denotes the strength of the effective spin–orbit coupling renormalized by many-body effects, and $p_\perp$ is the momentum of a conduction electron in the $k_x$–$k_y$ plane. $2|\alpha^* p_\perp|$ can be estimated from equation (2):[20,21,31]

$$|F_+ - F_-| = \frac{2c}{\hbar e}|\alpha^* p_\perp| m^*, \quad (2)$$

where $F_+$ and $F_-$ are the cross sections of a pair of spin-split Fermi surfaces at $k_z = 0$, $c$ is the speed of light, $\hbar$ is the Dirac constant, and $e$ is the elementary charge. We apply this equation to the spin splitting of $Cd_2Re_2O_7$ in the $k_x$–$k_y$ plane to estimate the magnitude of the ASOC; although the ASOC of $Cd_2Re_2O_7$ is not of the Rashba type, it gives an identical spin splitting in the $k_x$–$k_y$ plane. Provided that the two peaks with the dHvA frequencies of 2160 T and 1880 T (μ and λ branches) at $B$ // [001] are derived from the #273 and #274 Fermi surfaces, respectively, we obtain $2|\alpha^* p_\perp|$ = 67 K. Two additional assumptions made are that they come from only one of the tetragonal domains with $B$ along the tetragonal $c$ axis and that the effective mass of the latter equals to that of the former ($5.5 m_0$). Nevertheless, a comparable spin splitting should occur for the pair of #275 and #276; however, the identification of their dHvA signals remains ambiguous. In addition, for the pair of #271 and #272, equation (2) is not applicable as they are placed near the zone boundary.

The estimated magnitude of the ASOC for $Cd_2Re_2O_7$ is smaller than a noncentrosymmetric metal $CeRhSi_3$ ($2|\alpha^* p_\perp|$ = 140 K),[21] and is much smaller than those of other noncentrosymmetric metals: $LaRhSi_3$ (250 K),[20] $LaIrSi_3$ (1100 K)[20], and BiTeI (4600 K).[22] One of the reasons for the relatively small magnitude of the ASOC for $Cd_2Re_2O_7$ is likely that, in $Cd_2Re_2O_7$, the crystalline distortion accompanied by the ISB transition due to the electronic instability of SOCM is small to generate a small effective electric field. In contrast, BiTeI, for example, has a built-in polar axis with a large effective electric field in its original crystal structure, which should induce a much larger ASOC. Nevertheless, the small ASOC of $CeRhSi_3$ compared with that of $LaRhSi_3$ is ascribed to the difference in $m^*$; $m^*/m_0$ = 1.5 and 7–11 for $LaRhSi_3$ and $CeRhSi_3$, respectively.[21] This mass renormalization effect may be another reason for the small ASOC of $Cd_2Re_2O_7$ with $m^*$ = 5–9$m_0$.

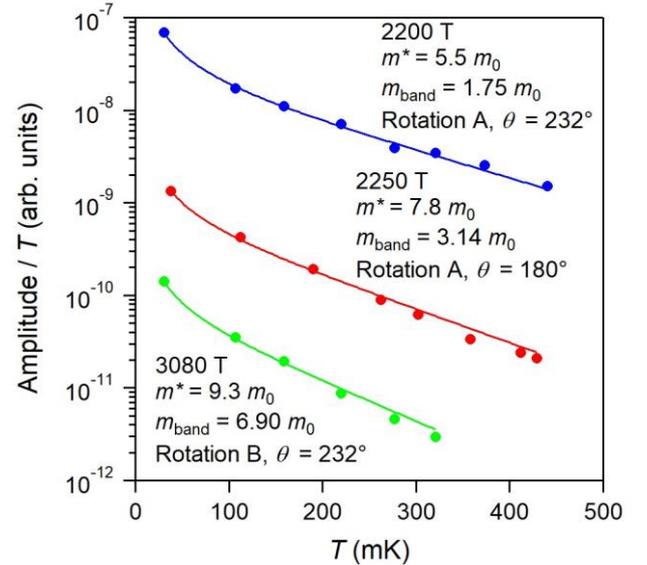

**Fig. 4.** (Color online) Temperature dependencies of the amplitudes of three dHvA oscillations at 2200 T and $B$ // [001], 2250 T, and $B$ // [111] in rotation A, and 3080 T and $\theta$ = 232° in rotation B. The former may correspond to the hole Fermi surface and the latter two to the electron Fermi surfaces. Each line on the data points represents a fitting to eq. (1). Effective masses estimated from the temperature dampings are $5.5m_0$, $7.8m_0$, and $9.3m_0$, which are larger than the corresponding band masses of $1.75m_0$, $3.14m_0$, and $6.90m_0$, respectively.

In conclusion, we have observed, for the first time, dHvA oscillations in the SOCM candidate $Cd_2Re_2O_7$. Spin-split electron Fermi surfaces are identified via comparison with a band structure calculation. The magnitude of the ASOC is estimated to be 67 K, which is much smaller than those of other noncentrosymmetric metals. This is possibly characteristic of the SOCM, in which ASOC is activated by the ISB associated with the Fermi liquid instability. In addition, we have observed large cyclotron masses of 5–9$m_0$.

**Acknowledgments** We thank J. Yamaura, H. Harima, T. Kobayashi, T. Hasegawa, M. Takigawa, T. Arima, Y. Motome for helpful comments and discussions. Y. M. is supported by the Materials Education Program for the Future Leaders in Research, Industry, and Technology (MERIT) created by the Ministry of Education, Culture, Sports, Science, and Technology of Japan (MEXT). This work was partially supported by KAKENHI Grant No. JP17H07349 and the Core-to-Core Program for Advanced Research Networks given by the Japan Society for the Promotion of Science (JSPS).

*E-mail: hiroi@issp.u-tokyo.ac.jp